\documentclass{elsart}
\usepackage{graphicx}
\begin{document}
\begin{frontmatter}
\title{Andreev Bound States in Ferromagnet - Superconductor Nanostructures
\thanksref{CM}}
\thanks[CM]{This work has been supported by Computational Magnetoelectronics 
            Research Training Network under Contract No. HPRN-CT-2000-00143}
\author{M. Krawiec \corauthref{cor}}
\author{, B. L. Gy\"{o}rffy and J. F. Annett}
\corauth[cor]{e-mail address: m.a.krawiec@bristol.ac.uk}
\address{H. H. Wills Physics Laboratory, University of Bristol, \\
         Tyndall Avenue, Bristol BS8 1TL, UK}

\begin{abstract}
We discuss the properties of a ferromagnet - superconductor heterostructure on
the basis of a Hubbard model featuring exchange splitting in the ferromagnet
and electron - electron attraction in the superconductor. We have solved the 
spin - polarized Hartree - Fock - Gorkov equations together with the Maxwell's 
equation (Ampere's law) fully self-consistently. We have found that a Proximity 
Effect - Fulde - Ferrell - Larkin - Ovchinnikov state is realized in such a 
heterostructure. It manifests itself in an oscillatory behavior of the pairing 
amplitude in the ferromagnet and spontaneously generated spin polarized current 
in the ground state. We argue that it is built up from the Andreev bound states,
whose energy can be tuned by the exchange splitting and hence can coincide with 
the Fermi energy giving rise to a current carrying $\pi$-state. We also suggest 
experiments to verify these predictions.
\end{abstract}
\end{frontmatter}


\section{\label{sec1} Introduction}

When a non-magnetic normal metal is in contact with a superconductor it 
acquires superconducting properties. This effect, known as the proximity effect 
\cite{Lambert}, has extensively been studied for almost $40$ years, and is 
rather well understood in terms of Andreev reflections \cite{Andreev}. On the 
other hand the proximity effect between a ferromagnet ($FM$) and a 
superconductor ($SC$) is less understood. However recent advance in 
nanofabrication made it possible to produce high quality $FM$/$SC$ interfaces 
\cite{fm_sc} and hence there is much current interest in the study of the 
interplay between magnetism and superocnductivity in heterostructures involving
such surfaces \cite{Wong}-\cite{Kontos2}.

It is widely accepted that ferromagnetism is destructive for superconductivity.
Thus one expects that the proximity effect in a ferromagnet should be very 
short ranged. Indeed, it has been predicted \cite{deJong} that Andreev 
reflections are suppressed due to the fact that impinging electrons and 
reflected holes occupy bands with different spin orientations. Surprisingly, 
some of the experiments seem to be in contradiction with the short range nature 
of the proximity effect. For instance, it was found \cite{Wong} that the 
transition temperature $T_c$ of the $FM$/$SC$ multilayers oscillates as a 
function of the $FM$ thickness. This curious behavior has been attributed to 
the formation of an effective $\pi$-junction \cite{Radovic}.

In general a system is in the $\pi$-phase if the $SC$ order parameter changes 
its sign across the junction. The $\pi$-junction behavior has been extensively 
studied in connection with the high-$T_c$ superconductors \cite{pi_junction}, 
where $SC$ order parameter changes its sign under $\pi/2$ rotation. This has 
tremendous consequences as it leads to the zero energy Andreev states 
\cite{Hu}, zero bias conductance peaks, paramagnetic Meissner effect and 
spontaneously generated currents.

Some other experiments show oscillatory behavior of the $T_c$, even though a 
$\pi$-junction cannot be realized in the geometry investigated. The example is 
a $FM$/$SC$/$FM$ trilayer \cite{Muhge}, where only one $SC$ layer is present. 
Recently oscillations of the density of states with $FM$ thickness in a 
$FM$/$SC$ bilayer has been also observed \cite{Kontos1}. Evidently a
conventional $\pi$-junction is also impossible in this case. Interestingly, 
such unusual behavior can be explained in terms of a Fulde - Ferrell - Larkin - 
Ovchinnikov ($FFLO$) - like state \cite{FFLO}, forming in the proximity
conditions.

Usually, when the exchange field is increased, one would expect that either the
field is too weak to break Cooper pairs or it leads through first order phase
transition to the normal state. However as it was noted in \cite{FFLO}, for 
certain values of the exchange field a new superconducting state can be 
realized. This $FFLO$ state features a spatially dependent order parameter 
and the current flow in the ground state. The total current consists of two 
contributions: one, which is due to the normal unpaired electrons and the 
other one, which is a supercurrent. These two parts cancel each other, so the 
Bloch theorem: no current in the ground state, is satisfied.

Similarly in the $FM$/$SC$ heterostructures: the oscillations of the pairing
amplitude have been predicted \cite{Buzdin}-\cite{Halterman} as well as
spontaneously generated current in the ground state \cite{KGA}. These features
give a strong evidence that the $FFLO$ state is really realized in $FM$/$SC$
nanostructures.

In the present paper we attempt to provide further insights into the physics of 
$FFLO$ state in $FM$/$SC$ heterostructures. In particular we investigate the 
properties of the Andreev bound states, pairing amplitude and the spontaneous 
current when the temperature is varied. We suggest that  temperature 
measurements of various  experimentally accessible quantities can contribute
much to the understanding of the superconductivity in $FM$/$SC$ bilayers. The 
paper is organized as follows: In the section  \ref{sec2} we introduce a simple 
model which can handle the main properties of $FM$/$SC$ systems. Some technical 
details concerning numerical implementation can be found in a subsequent paper 
\cite{KGA}. In sec. \ref{sec3} the nature of the Andreev bound states is 
discussed. We also show calculated temperature dependences of various 
quantities characterizing our system. In particular the density of states, 
which can be measured experimentally, can unambiguously confirm the current 
flowing in the ground state. Finally, the conclusions are given in sec. 
\ref{sec4}.


\section{\label{sec2} The model and the formal structure of the theory}

To study the properties of $FM$/$SC$ system we have adopted the $2D$ Hubbard
model featuring the exchange splitting in the ferromagnet and an electron -
electron attraction in superconductor. The Hamiltonian is:
\begin{eqnarray}
 H = 
 \sum_{ij\sigma} \left[ t_{ij} + \left( \frac{1}{2} E_{ex} \sigma - \mu \right) 
 \delta_{ij} \right] c^+_{i\sigma} c_{j\sigma} + 
 \frac{1}{2} \sum_{i\sigma} U_i n_{i\sigma} n_{i-\sigma}
 \label{Hamiltonian}
\end{eqnarray}
where in the presence of a vector potential $\vec{A}(\vec{r})$, the hopping
integral is given by $t_{ij} = - t e^{-i e \int_{\vec{r}_i}^{\vec{r}_{j}}
\vec{A}(\vec{r}) \cdot d\vec{r}}$ for nearest neighbor lattice sites, whose
positions are $\vec{r}_i$ and $\vec{r}_j$, and zero otherwise. The exchange
splitting $E_{ex}$ is only non-zero on the $FM$ side, unlike as $U_i$ (electron
- electron attraction) being non-zero only in $SC$. $\mu$ is the chemical
potential, $c^+_{i\sigma}$, ($c_{i\sigma}$) are the usual electron
creation (annihilation) operators and
$\hat n_{i\sigma} = c^+_{i\sigma} c_{i\sigma}$.

In the following we shall work within Spin - Polarized - Hartree - Fock -
Gorkov ($SPHFG$) approximation \cite{KGA} assuming periodicity in the direction
parallel to the interface while working in a real space in the direction
perpendicular. Labelling the layers by integer $n$ and $m$ at each $k_y$ point
of the Brillouin zone we shall solve the following $SPHFG$ equation:
\begin{eqnarray}
 \sum_{m',\gamma,k_y} H^{\alpha\gamma}_{nm'}(\omega,k_y)
 G^{\gamma\beta}_{m'm}(\omega,k_y) =
 \delta_{nm} \delta_{\alpha\beta}
 \label{HFG}
\end{eqnarray}
where the only non-zero elements are:
$H^{11}_{nm}$ and $H^{22}_{nm} = (\omega - \frac{1}{2} \sigma E_{ex} \pm \mu
 \pm  t cos(k_y \mp eA(n)))\delta_{nm} \pm t \delta_{n,n+1}$ for the upper and
lower sign respectively,
$H^{33}_{nm} = H^{11}_{nm}$ and $H^{44}_{nm} = H^{22}_{nm}$ with $\sigma$
replaced by $-\sigma$ and
$H^{12}_{nm} = H^{21}_{nm} = - H^{34}_{nm} = - H^{43}_{nm} =
\Delta_n \delta_{nm}$
and $G^{\alpha\beta}_{nm}$ is corresponding Green's function ($GF$).

As usual, the self-consistency is assured by the relations determining the $FM$
($m_n$) and $SC$ ($\Delta_n$) order parameters, current 
($J_{y\uparrow (\downarrow)}(n)$) and the vector potential ($A_y(n)$)
respectively:
\begin{eqnarray}
 m_n = n_{n\uparrow} - n_{n\downarrow} =
 \frac{2}{\beta}
 \sum_{ky} \sum^{2N-1}_{\nu = 0}
 {\rm Re} \left\{
 (G^{11}_{nn}(\omega_{\nu},k_y) - G^{33}_{nn}(\omega_{\nu},k_y))
 e^{(2 \nu + 1) \pi i / 2 N}
 \right\}
 \label{m}
\end{eqnarray}
\begin{eqnarray}
 \Delta_n = U_n \sum_{k_y}
 \langle c_{n\downarrow}(k_y) c_{n\uparrow}(k_y) \rangle =
 \frac{2 U_n}{\beta}
 \sum_{ky} \sum^{2N-1}_{\nu = 0}
 {\rm Re} \left\{
 G^{12}_{nn}(\omega_{\nu},k_y) e^{(2 \nu + 1) \pi i / 2 N}
 \right\}
 \label{Delta}
\end{eqnarray}
\begin{eqnarray}
 J_{y\uparrow (\downarrow)}(n) =
 \frac{4 e t}{\beta} \sum_{k_y} sin(k_y - e A_y(n))
 \sum^{2N-1}_{\nu = 0}
 {\rm Re} \left\{
 G^{11(33)}_{nn}(\omega_{\nu},k_y) e^{(2 \nu + 1) \pi i / 2 N}
 \right\}
 \label{current}
\end{eqnarray}
\begin{eqnarray}
 A_y(n+1) - 2 A_y(n) + A_y(n-1) = - 4 \pi J_y(n)
 \label{Maxwell}
\end{eqnarray}
The details of the calculations can be found in \cite{KGA}.


\section{\label{sec3} Results and discussion}

Since we have determined the $FM$ (\ref{m}) and the $SC$ (\ref{Delta}) order
parameters on both sides of the interface fully self-consistently, we were able
to study both $FM$ and $SC$ proximity effects. The $FM$ order parameter (spin
polarization) shows the usual Meissner - like behavior in the $SC$ \cite{KGA}, 
while the $SC$ pairing amplitude oscillates as we increase the thickness of 
the $FM$ slab \cite{Demler}-\cite{Halterman}. Similar oscillations are found if 
we fix the thickness of the $FM$ sample and change the exchange splitting 
\cite{KGA}. So this is the first confirmation of the $FFLO$ state in $FM$/$SC$ 
heterostructures \cite{Demler,Kontos1}.

\begin{figure}
 \begin{center}
  \resizebox{12cm}{!}{\includegraphics{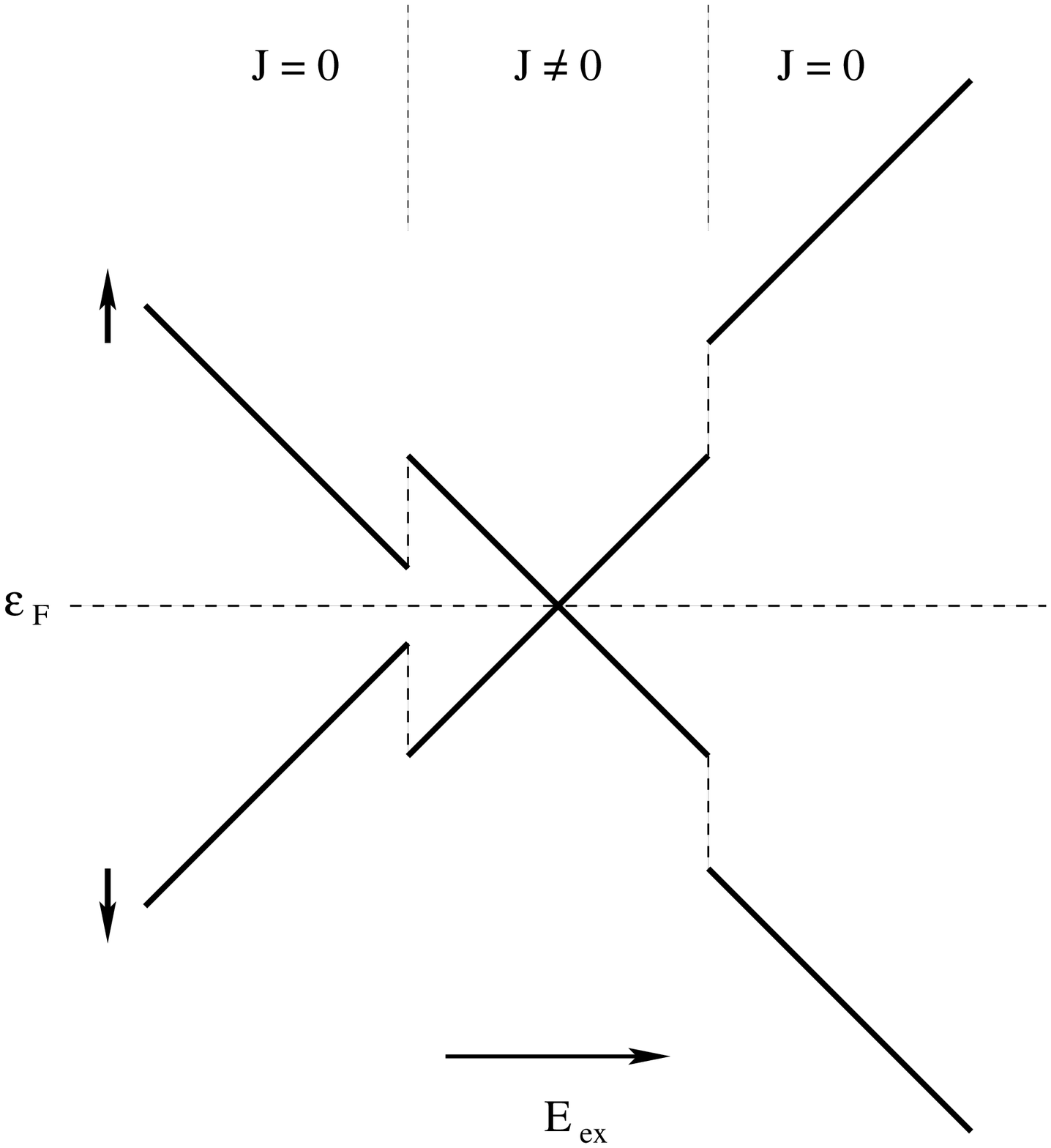}}
 \end{center} 
 \caption{Additional (Doppler) splitting of the Andreev bound state due to the
          current flowing in the $FM$/$SC$ heterostructure.}
 \label{f1}
\end{figure}

The interesting physics of such proximity structures is the formation of the
Andreev bound states \cite{deGennes}. They are `particle in a box' like states
due to the finite thickness of the $FM$ and have been discussed extensively
\cite{Vecino,KGA}. For a normal metal in contact with superconductor these
states are symmetrically located with the respect to the Fermi energy
$\varepsilon_F$. When the normal metal is replaced by ferromagnet the position
of these states is shifted due to the exchange splitting in $FM$
\cite{Andreev_position,Vecino,KGA}. This gives a possibility to shift such 
state to the zero energy ($\varepsilon_F$). Of course we should talk about 
Andreev bands rather than single states because in the $2D$ system we are
studying, there are the Andreev reflections for different angles with respect 
to the normal to the interface. In such a situation when the pairing amplitude 
at the end of the $FM$ slab ($FM$/vacuum interface) is equal to zero, 
spontaneous current is generated \cite{KGA}. The current flowing in the system 
produces a magnetic field, which splits this zero energy state thereby lowering 
the total energy of the system. If we follow the position of one particular 
state, forming an Andreev band, as the exchange splitting is increased, there 
is an additional (Doppler) shift when the current flows. The situation is 
schematically shown in the Fig. \ref{f1}. In $2$ or $3D$ geometry the whole 
Andreev band is split. The energy of such Doppler splitting is determined by 
the vector potential and in our model is given by 
$\delta \approx 2 e t \bar{A}_y$, where the layer averaged vector potential is
given by $\bar{A}_y = \Sigma_{n \in FM} A_y(n)/N_F$ for $N_F$ layers.

\begin{figure}
 \begin{center}
  \resizebox{15cm}{!}{\includegraphics{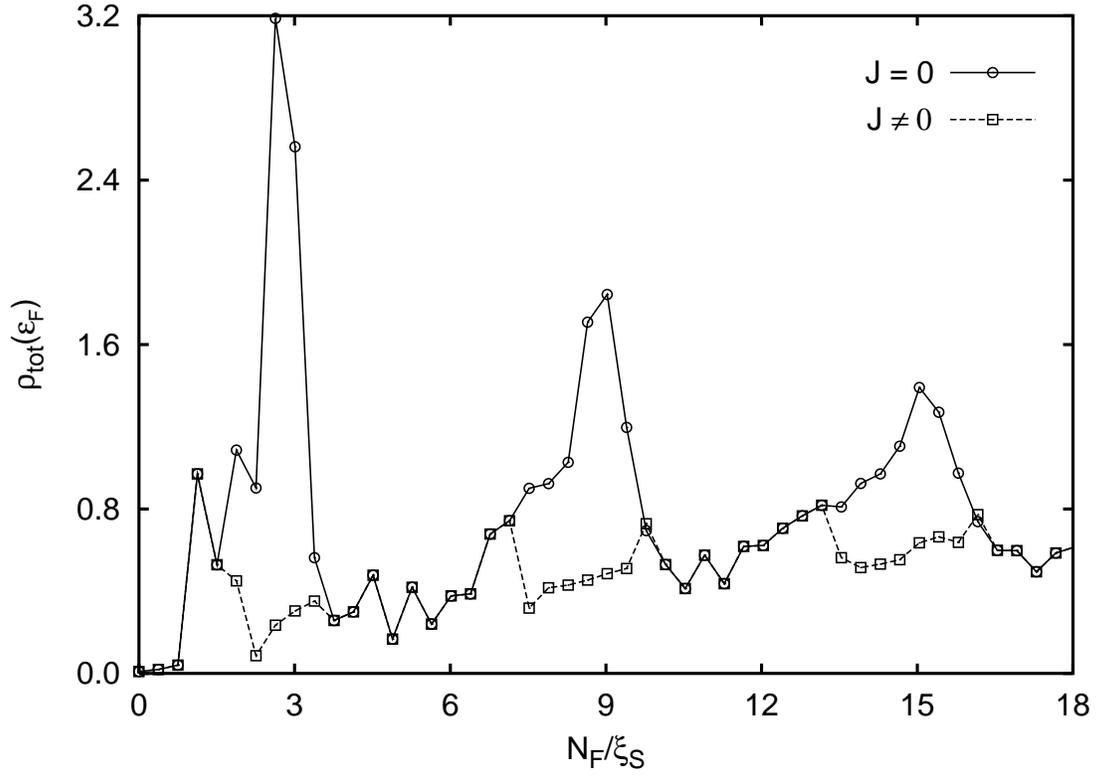}}
 \end{center} 
 \caption{The surface ($FM$/vacuum) density of states at the Fermi energy vs.
          number of the $FM$ layers. The squares (circles) correspond to the 
	  solution without (with) the current.}
 \label{f2}
\end{figure}

\begin{figure}
 \begin{center}
  \resizebox{15cm}{!}{\includegraphics{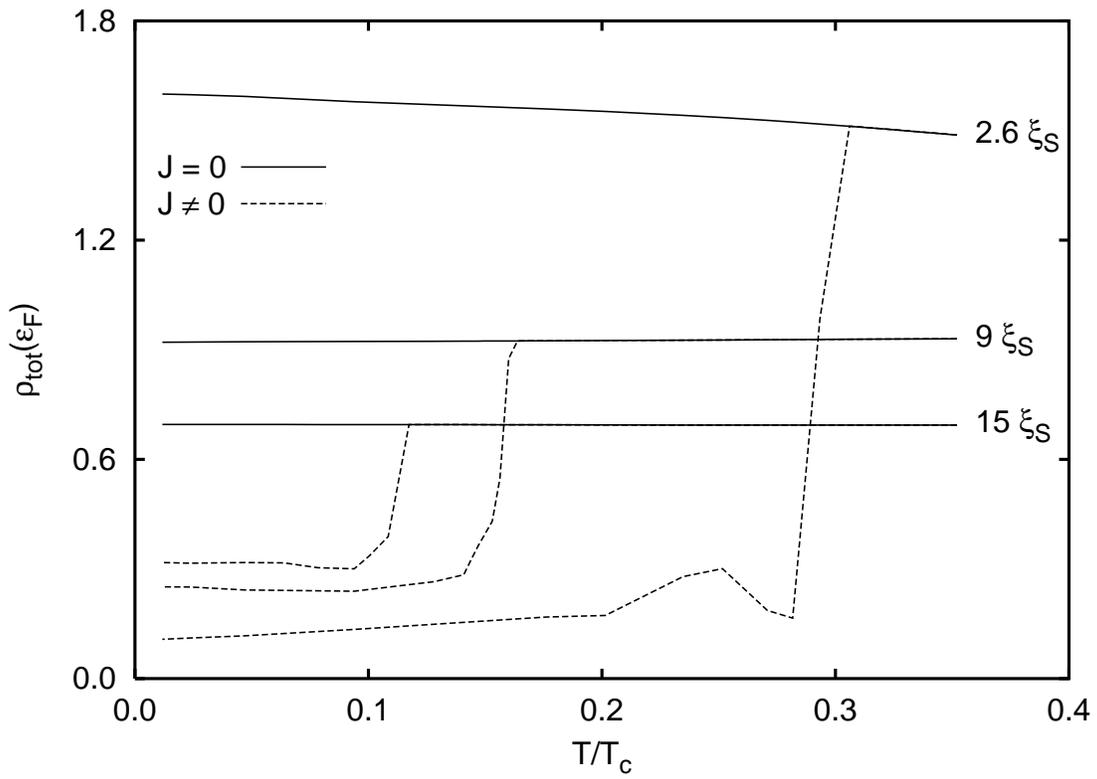}}
 \end{center} 
 \caption{The temperature dependence of the surface ($FM$/vacuum) density of
          states at the Fermi energy for various thicknesses of the $FM$ slab
	  in the figure. The solid (dashed) line corresponds to the solution 
	  without (with) the current.}
 \label{f3}
\end{figure}

The splitting of the Andreev bands due to the current flowing can be seen if we
plot the surface ($FM$/vacuum) density of states at the Fermi energy 
($\rho_{tot}(\varepsilon_F)$). This quantity can be directly measured 
experimentally using planar tunneling spectroscopy \cite{Kontos1}. The example 
of $\rho_{tot}(\varepsilon_F)$ is shown in the Fig. \ref{f2}. There is a
dramatic difference between solution with or without the current (in the later 
case the spontaneous current is constrained to be zero). In calculations it is
readily ascertained but experimentally it would be very difficult to judge if 
there is a current or not. However a plot of the temperature dependence of the 
$\rho_{tot}(\varepsilon_F)$ for fixed thickness clearly delineates this 
difference. At the thicknesses of $FM$ for which the current flows there is a 
huge drop in the $\rho_{tot}(\varepsilon_F)$ at characteristic temperature 
$T^{\ast} \approx (\xi_S/\lambda) T_c$, where $\xi_S$ and $\lambda$ are 
coherence length and penetration depth respectively. $T^{\ast}$ simply 
indicates the temperature at which magnetic instability, which leads to the 
generation of the current, takes the place. Such behavior is depicted in the
Fig. \ref{f3} and should be observable experimentally. If there is no current
the $DoS$ is due to the Andreev band and is almost constant (we are well below 
$T_c$), and as soon as the current starts to flow the Andreev band splits so we
observe a drop in $\rho_{tot}(\varepsilon_F)$. The important point is that 
$T^{\ast}$ and $T_c$ are different temperatures.

\begin{figure}
 \begin{center}
  \resizebox{15cm}{!}{\includegraphics{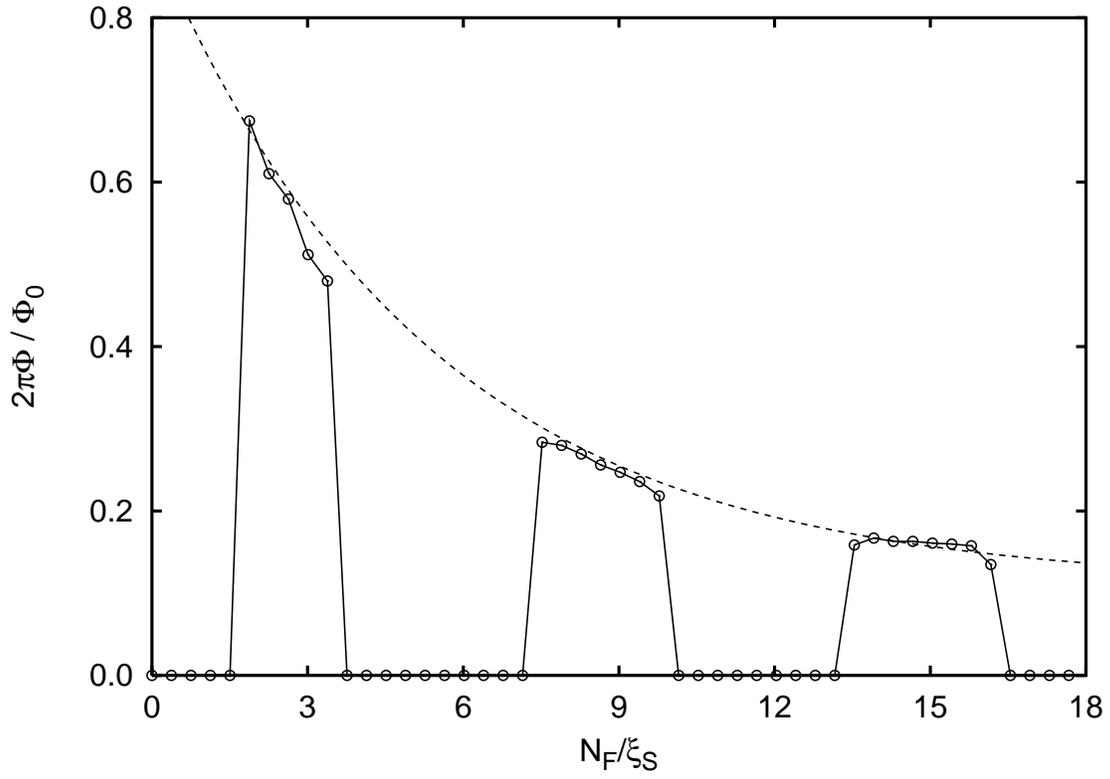}}
 \end{center} 
 \caption{The total magnetic flux per unit area in the $y$ direction vs. number
          of the $FM$ layers for $E_{ex}/\Delta_S = 0.5$. $\Phi_0 = h/2e$ is 
	  the flux quantum and $a$ is the lattice constant.}
 \label{f4}
\end{figure}

\begin{figure}
 \begin{center}
  \resizebox{15cm}{!}{\includegraphics{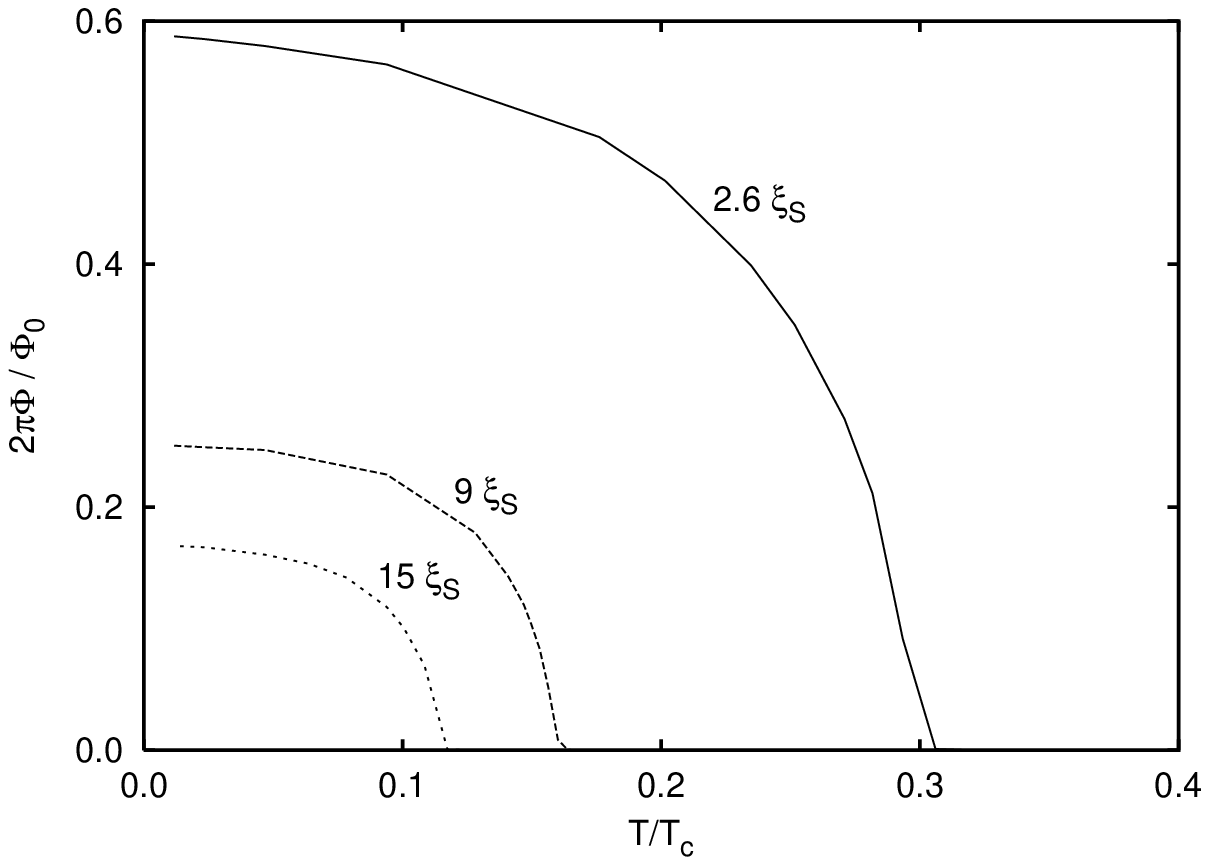}}
 \end{center} 
 \caption{The temperature dependence of the total magnetic flux for thickness 
          of the $FM$ slab $L/\xi_S = 2.6$ (solid), $6$ (dashed) and $15$ 
	  (dotted curve).}
 \label{f5}
\end{figure}

As we already mentioned, the spontaneous current generates the magnetic field,
which can also be measured experimentally by $SQUID$ techniques. Such 
spontaneous magnetic flux per unit area in the $y$ direction, parallel to the
interface, is defined by 
$\Phi = \Sigma_n \Phi(n) = \Sigma_n (A_y(n+1) - A_y(n))$ and is shown in the 
Fig. \ref{f4} as a function of the thickness of the $FM$ sample. We see that 
$\Phi$ is larger when $FM$ thickness is smaller and is of order of 
$10^{-2} - 10^{-1} \; \Phi_0$, where $\Phi_0 = h/2e$ is the flux quantum. 
Moreover it seems to show an exponential decay with increasing of the number 
of $FM$ layers. Similarly, in this case temperature measurement can provide 
information on the existence of the spontaneous current. Fig. \ref{f5} shows 
spontaneous magnetic flux for a number of thicknesses of the $FM$ slab. It is 
worthwhile to note that the behavior of $\Phi$ recalls the temperature 
dependence of the $SC$ order parameter in the $BCS$ theory.

Before closing the present discussion the following remark concerning the 
ferromagnet itself is in order. In general we should take into account the 
effect of the magnetic field coming from the localized moments of ferromagnet. 
These produce magnetic flux as well as surface currents. This effect could 
be important when the magnetization is perpendicular to the interface, as in 
our case. However if we deal with the weak ferromagnet or the sample has short 
lattice constant, the magnetic flux generated by local moments is much smaller 
than spontaneous one, and we can neglect this effect. In the present 
calculations we have taken the exchange splitting $E_{ex} = \Delta_S/2$, so 
this effect is of minor importance.


\section{\label{sec4} Conclusions}

In conclusion we have studied properties of the ferromagnet - superconductor
bilayer. We have shown that such a structure supports Andreev bound states 
forming Andreev bands, the position of which can be tuned by thickness of the
$FM$ sample. When a band crosses the Fermi energy, spontaneous current and 
magnetic flux is generated. We have found that the state with the current
flowing can be experimentally detected measuring the temperature dependence of
the density of states or the magnetic flux. We have argued that such current 
carrying state is a realization of the $FFLO$ state in $FM$/$SC$ proximity
system.



\newpage
%



\begin{thebibliography}{99}

\bibitem{Lambert} C. J. Lambert and R. Raimondi, J. Phys. Condens. Matter 
                  {\bf 10}, 901 (1998).
		  
\bibitem{Andreev} A. F. Andreev, Sov. Phys. JETP {\bf 19}, 1228 (1964).  

\bibitem{fm_sc} P. M. Tedrow and R. Meservey, Phys. Rep. {\bf 238}, 173 (1994); 
                V. T. Petrashov {\it et al.}, JETP Lett. {\bf 59}, 551 (1994); 
		V. T. Petrashov {\it et al.}, J. Low. Temp. Phys. {\bf 118}, 
		689 (2000); 
		P. Przyslupski {\it et al.}, Czech. J. Phys. {\bf 46}, 1355
		(1996); 
		H. -U. Habermeier {\it et al.}, Physica {\bf C364-365}, 298
		(2001).

\bibitem{Wong} H. K. Wong {\it et al.}, J. Low. Temp. Phys. {\bf 63}, 307
               (1986); 
	       J. S. Jiang {\it et al.}, Phys. Rev. Lett. {\bf 74}, 314 (1995).

\bibitem{Strunk} C. Strunk {\it et al.}, Phys. Rev. {\bf B49}, 4053 (1994); 
                 J. Aarts {\it et al.}, Phys. Rev. {\bf B56}, 2779 (1997)

\bibitem{Muhge} Th. M\"{u}hge {\it et al.}, Phys. Rev. Lett. {\bf 77}, 1857
	        (1996).

\bibitem{Kontos1} T. Kontos {\it et al.}, Phys. Rev. Lett. {\bf 86}, 304 (2001).


\bibitem{Ryazanov} V. V. Ryazanov {\it et al.}, Physica C {\bf 341-348}, 1613
                   (2000); 
                   Phys. Rev. Lett. {\bf 86}, 2427
                   (2001).
		   
\bibitem{Kontos2} T. Kontos {\it et al.}, Phys. Rev. Lett. {\bf 89}, 137007 
                  (2002)

\bibitem{deJong} M. J. M. de Jong and C. W. J. Beenakker, Phys. Rev. Lett. 
                 {\bf 74}, 1657 (1995).

\bibitem{Radovic} Z. Radovi\'{c} {\it et al.}, Phys. Rev. {\bf B44}, 759 
                  (1991). 
\bibitem{pi_junction} S. Kashiwaya and Y. Tanaka, Rep. Prog. Phys. {\bf 63}, 
                      1641 (2000);
                      T. L\"{o}fwander {\it et al.}, Supercond. Sci. Technol. 
		      {\bf 14}, R53 (2001).

\bibitem{Hu} C. Hu, Phys. Rev. Lett. {\bf 72}, 1526 (1994).

\bibitem{FFLO} P. Fulde and A. Ferrell, Phys. Rev. {\bf 135}, A550 (1964); 
               A. Larkin and Y. Ovchinnikov, Sov. Phys. JETP {\bf 20}, 762
	       (1965). 

\bibitem{Buzdin} A. I. Buzdin {\it et al.}, JETP Lett. {\bf 35}, 178 (1982); 
                 A. I. Buzdin and M. V. Kuprianov, JETP Lett. {\bf 52}, 487 
		 (1990).

\bibitem{Demler} E. A. Demler {\it et al.}, Phys. Rev. {\bf B55}, 15 174 (1997).

\bibitem{Vecino} E. Vecino {\it et al.}, Phys. Rev. {\bf B64} 184502 (2001).

\bibitem{Halterman} K. Halterman and O. T. Valls, Phys. Rev. {\bf 65} 014509
                    (2002);
                    preprint cond-mat/0205518.

\bibitem{KGA} M. Krawiec {\it et al.}, Phys. Rev. {\bf B66}, 172505 (2002); 
              cond-mat/0207135.

\bibitem{deGennes} P. G. de Gennes and D. Saint-James, Phys. Lett. {\bf 4}, 151
                   (1963).

\bibitem{Andreev_position} S. V. Kuplevakhskii and I. I. Falko, JETP Lett. 
                           {\bf 52}, 340 (1990); 
			   M. Zareyan {\it et al.}, Phys. Rev. Lett. {\bf 86},
			   308 (2001).
\end{thebibliography}
\end{document}